\documentclass
[notitlepage,groupedaddress,secnumarabic,prd,showpacs,showkeys,11pt]{revtex4}%
\usepackage{amssymb}
\usepackage{amsfonts}
\usepackage{amsmath}
\usepackage{latexsym}
\usepackage{graphicx}
\usepackage{appendix}%
\begin{document}
\title{Consistency of equations of motion in conformal frames}
\author{J.R. Morris}
\affiliation{Physics Dept., Indiana University Northwest, 3400 Broadway, Gary, Indiana,
46408, USA\ \ }

\begin{abstract}
Four dimensional scalar-tensor theory is considered within two conformal
frames, the Jordan frame (JF) and the Einstein frame (EF). The actions for the
theory are equivalent and equations of motion can be obtained from each
action. It is found that the JF equations of motion, expressed in terms of EF
variables, translate directly into and agree with the EF equations of motion
obtained from the EF action, provided that certain simple consistency
conditions are satisfied, which is always the case. The implication is that a
solution set obtained in one conformal frame can be reliably translated into a
solution set for the other frame, and therefore the two frames are, at least,
mathematically equivalent.

\end{abstract}

\maketitle

\section{Introduction}

\ \ The \textquotedblleft frames issue\textquotedblright\ has long been in
existence and has stirred much debate as to whether the Jordan frame and
Einstein frames are mathematically and physically equivalent, i.e., simply two
equivalent representations of the same theory or not. (See, for example,
\cite{Faraoni1}-\cite{Crooks} and references therein.) In the Jordan frame
(JF) of a scalar-tensor theory a \textquotedblleft dilaton\textquotedblright%
\ scalar field, say $\varphi$, couples nonminimally to the Ricci scalar
$\tilde{R}[\tilde{g}_{\mu\nu}]$ in the action, whereas a conformal
transformation to the Einstein frame (EF) results in an action where the
dilaton decouples from the new Ricci scalar $R[g_{\mu\nu}]$, but becomes
coupled to the matter sector \cite{FMbook},\cite{Damour}. In addition, the
lagrangian for the dilaton changes form under the conformal transformation
(see, e.g., \cite{Faraoni1}-\cite{Kaiser}). The classic 1962 paper by Dicke
\cite{Dicke62}, for example, convincingly argues that, at least at the
classical level, these are simply two different representations of the same
theory. This conclusion is based upon the \textit{equivalence of the actions}
in the two different frames. This point of view has often been exploited to
choose a convenient frame where an analysis of the system can be done, with
the reasonable assumption that the solutions of the equations of motion can be
reliably translated into the other frame. However, a physical equivalence of
the two frames has often been questioned or challenged (see, e.g.,
\cite{Faraoni2},\cite{Bhadra}).

\bigskip

\ \ Here, we take a look at a four dimensional scalar-tensor theory of a
fairly general form, similar to that used by Flanagan \cite{Flanagan} and
Damour and Esposito-Far\`{e}se \cite{Damour}, and consider the action in both
frames. The JF and EF actions are equivalent, up to discarded surface terms.
From each action the equations of motion can be obtained. However, what is new
here is that we then \textit{use the same conformal transformation to
translate one set of equations of motion into the variables of the other
frame}. For example, we take the JF equations of motion, expressed in terms of
the JF variables, say $\tilde{g}_{\mu\nu}$, $\tilde{R}[\tilde{g}_{\mu\nu}]$,
$\tilde{R}_{\mu\nu}$, and $\varphi$, and rewrite these JF equations of motion
in terms of the EF variables $g_{\mu\nu}$, $R[g_{\mu\nu}]$, $R_{\mu\nu}$, and
$\varphi$. A comparison is then made between the set of \textit{translated
equations of motion} and the \textit{action-based equations of motion}, both
expressed with the same variables. What we find is that these sets of
equations are the same, i.e., one set maps smoothly into the other, provided
that simple consistency conditions are satisfied, which, as another new
result, we show is always the case. So, along with the \textit{actions} for
the two frames being equivalent, up to discarded surface terms (i.e., the
lagrangians are equivalent up to discarded total divergences), we find that
the \textit{equations of motion} are also equivalent. It is felt that these
results further fortify the conclusion that the two frames are, at least,
mathematically equivalent. Therefore, mathematical solutions in one frame can
be reliably translated into the solutions of the other frame. Whether this
implies a physical equivalence of frames, as well, has been debated (e.g.,
\cite{Faraoni1},\cite{Faraoni2},\cite{Flanagan},\cite{Dicke62}).

\bigskip

\ \ The present study is motivated by occasional questions of whether the JF
and EF action-based equations of motion for a theory do in fact map into one
another under a conformal transformation. In other words, does an equivalence
of actions necessarily imply an equivalence of equations of motion in
different frames? After all, a conformal transformation of the JF Ricci scalar
term $\sqrt{\tilde{g}}\tilde{R}[\tilde{g}_{\mu\nu}]$ produces the EF term
$\sqrt{g}R[g_{\mu\nu}]$ in the action, along with kinetic terms for $\varphi$,
and total divergences in the lagrangian which are dropped. So, even though the
\textit{actions are equivalent}, one may wonder if the \textit{equations of
motion, and their solutions, are always equivalent}, given the mixing of the
metric and dilaton fields.

\bigskip

\ It is shown that the JF matter field equations, the JF Einstein equation,
and the JF dilaton field equation map into the corresponding EF equations,
provided that certain \textit{consistency conditions} are met. It is shown
that \textit{these conditions are always met}.

\section{Action-based equations of motion}

\ \ We assume a scalar-tensor theory action with the following forms (see,
e.g., \cite{JM01} for the case $\mathcal{\tilde{L}}_{m}=0$):
\begin{subequations}
\label{1}%
\begin{align}
S &  =\int d^{4}x\sqrt{\tilde{g}}\left\{  \frac{F(\varphi)}{2\kappa^{2}}%
\tilde{R}[\tilde{g}_{\mu\nu}]+\frac{1}{2}\tilde{k}(\varphi)\tilde{g}^{\mu\nu
}\partial_{\mu}\varphi\partial_{\nu}\varphi-V(\varphi)+\tilde{h}%
(\varphi)\mathcal{\tilde{L}}_{m}(\tilde{g}^{\mu\nu},\varphi,\psi)\right\}
\label{1a}\\
&  =\int d^{4}x\sqrt{g}\left\{  \frac{1}{2\kappa^{2}}R[g_{\mu\nu}]+\frac{1}%
{2}k(\varphi)g^{\mu\nu}\partial_{\mu}\varphi\partial_{\nu}\varphi
-U(\varphi)+h(\varphi)\mathcal{L}_{m}(g^{\mu\nu},\varphi,\psi)\right\}
\label{1b}%
\end{align}

where $\kappa=\sqrt{8\pi G}$ is the inverse of the reduced Planck mass,
$\mathcal{\tilde{L}}_{m}$ ($\mathcal{L}_{m}$) is the matter field lagrangian
in the JF (EF), and $\tilde{h}$ ($h$) is a dilatonic coupling function to
matter fields. The \textquotedblleft matter\textquotedblright\ fields are
those fields other than the metric or dilaton, and are represented
collectively by $\psi$. The terms $\tilde{h}\mathcal{\tilde{L}}_{m}$ and
$h\mathcal{L}_{m}$ are independent of $\partial_{\mu}\varphi$, but we allow
for the possibility that $\varphi$ couples to matter fields through a
potential $W(\varphi,\psi)$ in the lagrangians $\mathcal{\tilde{L}}_{m}$ and
$\mathcal{L}_{m}$. We also note that $\mathcal{\tilde{L}}_{m}$ and
$\mathcal{L}_{m}$ are the same lagrangians, but expressed with different
metrics for the kinetic terms. Various choices for the functions $F$,
$\tilde{k}$, $V$, and $\tilde{h}$ can accommodate different types of
scalar-tensor theory, including generalized Brans-Dicke theories and higher
dimensional theories dimensionally reduced to four dimensions. The form
(\ref{1a}) is the JF representation of the action, and (\ref{1b}) is the EF
representation. The conformal transformation connecting the two frames is
given by%
\end{subequations}
\begin{equation}
\tilde{g}_{\mu\nu}=\Omega^{2}g_{\mu\nu}=F^{-1}g_{\mu\nu},\ \ \ F(\varphi
)=e^{\varphi},\ \ \ \Omega(\varphi)=F^{-1/2}(\varphi)=e^{-\varphi/2}%
,\ \ \ \ln\Omega=-\frac{1}{2}\varphi\label{2}%
\end{equation}

where here we have chosen to parametrize $F$ by the exponential of a
dimensionless scalar field $\varphi=\ln F$. (We can relate $\varphi$ to a
scalar $\phi$ with canonical mass dimension by writing, e.g., $\varphi
=\kappa\phi$.) Note that in the JF action $\tilde{g}_{\mu\nu}$ and $\varphi$
are treated as independent fields, but in the EF action $g_{\mu\nu}$ and
$\varphi$ are treated as independent fields, and $\mathcal{L}_{m}(e^{\varphi
}g^{\mu\nu})=\mathcal{\tilde{L}}_{m}(\tilde{g}^{\mu\nu})$, so the kinetic
terms in the matter lagrangian depend on different metrics and have different
functional forms in the two frames.

\bigskip

\ We use a notation where $Q^{\prime}=\partial Q/\partial\varphi
=\partial_{\varphi}Q$, $Q^{\prime\prime}=\partial_{\varphi}^{2}Q$, etc. for
some function $Q(\varphi)$, and $(\partial\varphi)^{2}=\partial_{\mu}%
\varphi\partial^{\mu}\varphi$, and $(\tilde{\partial}\varphi)^{2}%
=\partial_{\mu}\varphi\tilde{\partial}^{\mu}\varphi=\tilde{g}^{\mu\nu}%
\partial_{\mu}\varphi\partial_{\nu}\varphi$. The JF functions $\tilde{k}$,
$V$, $\tilde{h}$, $\mathcal{\tilde{L}}_{m}$ are left arbitrary, with forms
determined by the particular action being considered, and the parametrization
$F=e^{\varphi}$. Therefore the EF functions $k$, $U$, $h$, and $\mathcal{L}%
_{m}$ are also arbitrary. The conformal transformation gives the functions
$k(\varphi)$ and $U(\varphi)$ of the EF action to be \cite{JM01}%

\begin{equation}
k=\frac{1}{\kappa^{2}}\left(  \frac{3F^{\prime2}}{2F^{2}}+\frac{\kappa
^{2}\tilde{k}}{F}\right)  ,\ \ \ \ \ U(\varphi)=\frac{V(\varphi)}%
{F^{2}(\varphi)} \label{3}%
\end{equation}

By (\ref{2}) we have $F^{\prime}=F^{\prime\prime}=F$ and we define the
functions $\tilde{C}(\varphi)=\kappa^{2}\tilde{k}(\varphi)$ and $C(\varphi
)=\kappa^{2}k(\varphi)$, so that (\ref{2}) and (\ref{3}) give%
\begin{equation}
\tilde{g}_{\mu\nu}=e^{-\varphi}g_{\mu\nu},\ \ \ \ \ \tilde{g}^{\mu\nu
}=e^{\varphi}g^{\mu\nu},\ \ \ \ \ C=\left(  \frac{3}{2}+\tilde{C}e^{-\varphi
}\right)  ,\ \ \ \ \ U(\varphi)=e^{-2\varphi}V(\varphi)\label{4}%
\end{equation}

\bigskip

\ \ We can write the action $S$ in the forms $S=\int d^{4}x\sqrt{\tilde{g}%
}\mathcal{\tilde{L}}=\int d^{4}x\sqrt{g}\mathcal{L}$ and look at the variation
of $S$ with respect to matter fields, the metric, and the scalar dilaton field
to obtain the action-based equations of motion. Since $\sqrt{\tilde{g}%
}\mathcal{\tilde{L}}=\sqrt{g}\mathcal{L}$ we can use the result%
\begin{equation}
\mathcal{\tilde{L}}=\frac{\sqrt{g}}{\sqrt{\tilde{g}}}\mathcal{L}%
=F^{2}\mathcal{L} \label{5}%
\end{equation}

where use has been made of (\ref{2}) with
\begin{equation}
\tilde{g}_{\mu\nu}=F^{-1}g_{\mu\nu},\ \ \tilde{g}^{\mu\nu}=Fg^{\mu\nu
},\ \ \sqrt{\tilde{g}}=F^{-2}\sqrt{g}\label{6}%
\end{equation}

\subsection{Matter fields}

\textbf{\ \ }The equation of motion (EoM) for a matter field $\psi$ is
obtained by varying $S$ with respect to $\psi$, i.e., $\delta_{\psi}S=0$. We
then obtain the JF and EF EoM given, respectively, by
\begin{subequations}
\label{7}%
\begin{align}
\tilde{\nabla}_{\mu}\left[  \frac{\partial\mathcal{\tilde{L}}}{\partial
(\partial_{\mu}\psi)}\right]  -\frac{\partial\mathcal{\tilde{L}}}{\partial
\psi} &  =0\label{7a}\\
\nabla_{\mu}\left[  \frac{\partial\mathcal{\mathcal{L}}}{\partial
(\partial_{\mu}\psi)}\right]  -\frac{\partial\mathcal{\mathcal{L}}}%
{\partial\psi} &  =0\label{7b}%
\end{align}

where $\tilde{\nabla}_{\mu}$ and $\nabla_{\mu}$ are covariant derivatives with
respect to $\tilde{g}_{\mu\nu}$ and $g_{\mu\nu}$, respectively, and
$\nabla_{\mu}V^{\mu}=\frac{1}{\sqrt{g}}\partial_{\mu}\left[  \sqrt{g}g^{\mu
\nu}V_{\nu}\right]  $ with a similar expression holding for $\tilde{\nabla
}_{\mu}\tilde{V}^{\mu}$.

\subsection{Dilaton and Einstein equations}

\ \ The Jordan frame equations of motion that follow from the action $S$ in
(\ref{1a}) due to variations with respect to $\tilde{g}^{\mu\nu}$ and
$\varphi$ are, for arbitrary\ functions, $\tilde{k}(\varphi)$, $V(\varphi)$,
and $\tilde{h}(\varphi)$, given by (see \cite{JM01}, for example, for the case
$\mathcal{\tilde{L}}_{m}\rightarrow0$)%
\end{subequations}
\begin{align}
\tilde{G}_{\mu\nu} &  =-\tilde{\nabla}_{\mu}\partial_{\nu}\varphi-\left(
1+\tilde{C}e^{-\varphi}\right)  \partial_{\mu}\varphi\partial_{\nu}%
\varphi\nonumber\\
&  \ \ \ \ \ \ +\tilde{g}_{\mu\nu}\left\{  \left(  1+\frac{1}{2}\tilde
{C}e^{-\varphi}\right)  (\tilde{\partial}\varphi)^{2}+\tilde{\square}%
\varphi-\kappa^{2}e^{-\varphi}V\right\}  -\kappa^{2}e^{-\varphi}\tilde
{h}\tilde{T}_{\mu\nu}\label{8}%
\end{align}

\begin{equation}
\tilde{C}\tilde{\square}\varphi+\frac{1}{2}\tilde{C}^{\prime}(\tilde{\partial
}\varphi)^{2}+\kappa^{2}V^{\prime}-\frac{1}{2}e^{\varphi}\tilde{R}-\kappa
^{2}\partial_{\varphi}(\tilde{h}\mathcal{\tilde{L}}_{m})=0 \label{9}%
\end{equation}

where $\tilde{G}_{\mu\nu}=\tilde{R}_{\mu\nu}-\frac{1}{2}\tilde{g}_{\mu\nu
}\tilde{R}$ is the JF Einstein tensor and we have used $F=e^{\varphi}$ along
with the definition $\tilde{C}=\kappa^{2}\tilde{k}$, and
\begin{equation}
\tilde{T}_{\mu\nu}=\frac{2}{\sqrt{\tilde{g}}}\frac{\partial(\sqrt{\tilde{g}%
}\mathcal{\tilde{L}}_{m})}{\partial\tilde{g}^{\mu\nu}},\ \ \ \ \ T_{\mu\nu
}=\frac{2}{\sqrt{g}}\frac{\partial(\sqrt{g}\mathcal{L}_{m})}{\partial
g^{\mu\nu}} \label{10}%
\end{equation}
is the stress-energy tensor of the matter in the JF and EF, respectively. Now,
(\ref{9}) can be rewritten by taking the trace of (\ref{8}), with $\tilde
{R}=-\tilde{G}=-\tilde{g}^{\mu\nu}\tilde{G}_{\mu\nu}$, and inserting this into
(\ref{9}). We then have the JF dilaton EoM%
\begin{equation}
\left(  \frac{3}{2}e^{\varphi}+\tilde{C}\right)  \tilde{\square}%
\varphi+\left[  \frac{3}{2}e^{\varphi}+\frac{1}{2}(\tilde{C}+\tilde{C}%
^{\prime})\right]  (\tilde{\partial}\varphi)^{2}+\kappa^{2}e^{2\varphi
}U^{\prime}-\kappa^{2}\left[  \partial_{\varphi}(\tilde{h}\mathcal{\tilde{L}%
}_{m})+\frac{1}{2}\tilde{h}\tilde{T}\right]  =0 \label{11}%
\end{equation}

where we have used the fact that $V^{\prime}-2V=e^{2\varphi}U^{\prime}$.

\bigskip

\ \ We can obtain the Einstein frame equations of motion from the action $S$
in (\ref{1b}) by making the replacements $F\rightarrow\hat{F}=1$, $F^{\prime
}\rightarrow\hat{F}^{\prime}=0$, $\tilde{k}\rightarrow k(\varphi)$, $\tilde
{C}\rightarrow C$, $\tilde{h}\rightarrow h$, $\mathcal{\tilde{L}}%
_{m}\rightarrow\mathcal{L}_{m}$, $\tilde{g}_{\mu\nu}\rightarrow g_{\mu\nu}$,
$\tilde{R}_{\mu\nu}\rightarrow R_{\mu\nu}$, $\tilde{R}\rightarrow R$, and
$V\rightarrow U$ in the equations of (\ref{8}) and (\ref{9}), where
$C(\varphi)$ and $U(\varphi)$ are given by (\ref{4}) with $C^{\prime
}=e^{-\varphi}(\tilde{C}^{\prime}-\tilde{C})$. Doing so, we obtain the EF
equations of motion:%
\begin{equation}
G_{\mu\nu}=-\left(  \frac{3}{2}+\tilde{C}e^{-\varphi}\right)  \partial_{\mu
}\varphi\partial_{\nu}\varphi+g_{\mu\nu}\left\{  \frac{1}{2}\left(  \frac
{3}{2}+\tilde{C}e^{-\varphi}\right)  (\partial\varphi)^{2}-\kappa
^{2}U\right\}  -\kappa^{2}hT_{\mu\nu} \label{12}%
\end{equation}

\begin{equation}
\left(  \frac{3}{2}+\tilde{C}e^{-\varphi}\right)  \square\varphi+\frac{1}%
{2}e^{-\varphi}\left(  \tilde{C}^{\prime}-\tilde{C}\right)  (\partial
\varphi)^{2}+\kappa^{2}U^{\prime}-\kappa^{2}\partial_{\varphi}(h\mathcal{L}%
_{m})=0 \label{13}%
\end{equation}

\bigskip

\ \ We now have the action-based Jordan frame equations of motion given by
(\ref{7a}), (\ref{8}), and (\ref{11}), derived from the action (\ref{1a}), and
the action-based Einstein frame equations of motion given by (\ref{7b}),
(\ref{12}), and (\ref{13}) derived from the action (\ref{1b}). The matter that
we now want to investigate is whether equations (\ref{7a}), (\ref{8}), and
(\ref{11}) will map directly into equations (\ref{7b}), (\ref{12}), and
(\ref{13}) under the conformal transformation.

\section{Translation and consistency of equations of motion}

\ \ We now wish to use the conformal transformation of (\ref{2}) and (\ref{6})
to rewrite the Jordan frame equations of motion in terms of Einstein frame
variables and then compare the translated Jordan frame equations to the
Einstein frame equations obtained from the Einstein frame action. We point out
that we use the signs and conventions of \cite{BDbook}, and our metric
$g_{\mu\nu}$ has opposite signature to that of \cite{Waldbook} and \cite{MTW},
and our Ricci tensor $R_{\mu\nu}$ is minus that of \cite{Waldbook} and
\cite{MTW}, while our Ricci scalar $R$ has the same sign as that of
\cite{Waldbook} and \cite{MTW}, with $R_{\mu\nu}=\partial_{\nu}\Gamma
_{\mu\lambda}^{\lambda}-\partial_{\lambda}\Gamma_{\mu\nu}^{\lambda}%
-\Gamma_{\mu\nu}^{\rho}\Gamma_{\rho\sigma}^{\sigma}+\Gamma_{\mu\sigma}^{\rho
}\Gamma_{\nu\rho}^{\sigma}$ and $R=g^{\mu\nu}R_{\mu\nu}$.

\bigskip

\ \ \textbf{Translation.}\ \ From (\ref{1}), with the use of (\ref{5}), we
find that the matter equation (\ref{7a}) directly yields (\ref{7b}). On the
other hand, some algebra (see Appendix) yields the translated form of
(\ref{11}) as%
\begin{equation}
\left(  \frac{3}{2}+\tilde{C}e^{-\varphi}\right)  \square\varphi+\frac{1}%
{2}e^{-\varphi}(\tilde{C}^{\prime}-\tilde{C})(\partial\varphi)^{2}+\kappa
^{2}U^{\prime}-\kappa^{2}e^{-2\varphi}\left[  \partial_{\varphi}(\tilde
{h}\mathcal{\tilde{L}}_{m})+\frac{1}{2}\tilde{h}\tilde{T}\right]  =0\label{14}%
\end{equation}

and the translated form of (\ref{8}) is%
\begin{equation}
G_{\mu\nu}=-\left(  \frac{3}{2}+\tilde{C}e^{-\varphi}\right)  \partial_{\mu
}\varphi\partial_{\nu}\varphi+\frac{1}{2}g_{\mu\nu}\left(  \frac{3}{2}%
+\tilde{C}e^{-\varphi}\right)  (\partial\varphi)^{2}-g_{\mu\nu}\kappa
^{2}U-\kappa^{2}e^{-\varphi}\tilde{h}\tilde{T}_{\mu\nu}\label{15}%
\end{equation}

These are the translated JF equation of motion for the dilaton $\varphi$ and
the translated JF Einstein equation written in terms of the EF metric
$g_{\mu\nu}$. Upon comparing (\ref{11}) to (\ref{14}) and (\ref{8}) to
(\ref{15}), we see that the EF equations of motion coincide with the
translated JF ones provided that%
\begin{equation}
\partial_{\varphi}(h\mathcal{L}_{m})=e^{-2\varphi}\left[  \partial_{\varphi
}(\tilde{h}\mathcal{\tilde{L}}_{m})+\frac{1}{2}\tilde{h}\tilde{T}\right]
\label{16}%
\end{equation}

and%
\begin{equation}
e^{-\varphi}\tilde{h}\tilde{T}_{\mu\nu}=hT_{\mu\nu}\label{17}%
\end{equation}

Therefore, (\ref{16}) and (\ref{17}) serve as consistency conditions for the
action-based JF and EF equations of motion to coincide. These conditions are
found to always be satisfied.

\bigskip

\ \ \textbf{Consistency.}\ \ Let us first look at (\ref{16}). Here we note
that in obtaining the JF EoM for $\varphi$ we treat $\varphi$ as independent
of $\tilde{g}_{\mu\nu}$, whereas in obtaining the EF EoM for $\varphi$ we
treat $\varphi$ as independent of $g_{\mu\nu}$. In other words, in evaluating
$\partial_{\varphi}\mathcal{\tilde{L}}_{m}=\mathcal{\tilde{L}}_{m}^{\prime}$
and $\partial_{\varphi}\mathcal{L}_{m}=\mathcal{L}_{m}^{\prime}$ we mean
\begin{subequations}
\label{18}%
\begin{align}
\mathcal{\tilde{L}}_{m}^{\prime}  & =\partial_{\varphi}%
\mathcal{\mathcal{\tilde{L}}}_{m}(\tilde{g}^{\mu\nu},\varphi)\big|_{\tilde{g}%
}=-W^{\prime}(\varphi)\label{18a}\\
\mathcal{L}_{m}^{\prime}  & =\partial_{\varphi}\mathcal{L}_{m}(e^{\varphi
}g^{\mu\nu},\varphi)\big|_{g}=\frac{\partial\mathcal{L}_{m}}{\partial\tilde
{g}^{\mu\nu}}\frac{\partial(e^{\varphi}g^{\mu\nu})}{\partial\varphi}%
-W^{\prime}(\varphi)=\tilde{g}^{\mu\nu}\frac{\partial\mathcal{L}_{m}}%
{\partial\tilde{g}^{\mu\nu}}-W^{\prime}\label{18b}%
\end{align}

where $\big|_{\tilde{g}}$ means evaluate at constant $\tilde{g}^{\mu\nu}$ and
$\big|_{g}$ means evaluate at constant $g^{\mu\nu}$. Now we can rewrite the
right hand side of (\ref{16}) in terms of EF variables. To do this, we compare
the dilaton-matter part of the action in (\ref{1a}) and (\ref{1b}) where we
see that $\sqrt{\tilde{g}}\tilde{h}=\sqrt{g}h$, using the fact that
$\mathcal{\tilde{L}}_{m}(\tilde{g}^{\mu\nu})=\mathcal{L}_{m}(e^{\varphi}%
g^{\mu\nu})$, and therefore%
\end{subequations}
\begin{equation}
\tilde{h}=\frac{\sqrt{g}}{\sqrt{\tilde{g}}}h=e^{2\varphi}h,\ \ \ \ \ \tilde
{h}\mathcal{\tilde{L}}_{m}=e^{2\varphi}h\mathcal{L}_{m}\label{19}%
\end{equation}

Furthermore, using $\sqrt{\tilde{g}}\tilde{h}\mathcal{\tilde{L}}_{m}(\tilde
{g}^{\mu\nu})=\sqrt{g}h\mathcal{L}_{m}(e^{\varphi}g^{\mu\nu})$ and the
definitions of $\tilde{T}_{\mu\nu}$ and $T_{\mu\nu}$ in (\ref{10}), it follows
that%
\begin{equation}
\tilde{h}\tilde{T}_{\mu\nu}=\frac{2\tilde{h}}{\sqrt{\tilde{g}}}\frac
{\partial(\sqrt{\tilde{g}}\mathcal{\tilde{L}}_{m})}{\partial\tilde{g}^{\mu\nu
}}=\frac{2he^{2\varphi}}{\sqrt{g}}\frac{\partial(\sqrt{g}\mathcal{L}_{m}%
)}{\partial\tilde{g}^{\mu\nu}}=\frac{2he^{\varphi}}{\sqrt{g}}\frac
{\partial(\sqrt{g}\mathcal{L}_{m})}{\partial g^{\mu\nu}}=e^{\varphi}hT_{\mu
\nu}\label{20}%
\end{equation}

We therefore have the results%
\begin{equation}
\tilde{h}\tilde{T}_{\mu\nu}=e^{\varphi}hT_{\mu\nu},\ \ \ \ \tilde{h}\tilde
{T}=e^{2\varphi}hT\label{21}%
\end{equation}

where $\tilde{g}^{\mu\nu}=e^{\varphi}g^{\mu\nu}$ has been used. We note that
(\ref{21}) guarantees that the consistency condition (\ref{17}) is satisfied.

\bigskip

\ \ Next, with the help of (\ref{18}), (\ref{19}), and (\ref{21}), we can
rewrite the consistency condition (\ref{16}) as%
\begin{equation}
\mathcal{\mathcal{L}}_{m}^{\prime}\big|_{g}=\left(  2\mathcal{L}_{m}+\frac
{1}{2}T\right)  +\mathcal{\tilde{L}}_{m}^{\prime}\big|_{\tilde{g}}%
\implies\tilde{g}^{\mu\nu}\frac{\partial\mathcal{L}_{m}}{\partial\tilde
{g}^{\mu\nu}}-W^{\prime}=\left(  2\mathcal{L}_{m}+\frac{1}{2}T\right)
-W^{\prime}(\varphi)\label{22}%
\end{equation}

where $\mathcal{L}_{m}^{\prime}=\partial_{\varphi}\mathcal{L}_{m}$ and
(\ref{18}) has been used. If the condition (\ref{22}) is satisfied, then the
EoM for $\varphi$ in the JF will be faithfully translated into the
action-based EF EoM for $\varphi$. Also note that this condition is
independent of the dilaton coupling functions $\tilde{h}$ and $h$, and depends
only upon the matter lagrangian and the trace of the stress-energy tensor
derived from it.

\bigskip

\ \ \ \ In fact, we can see that the condition (\ref{22}) holds, in general.
This follows from the definition of the stress-energy tensor (\ref{10}) along
with (\ref{19}) and (\ref{21}). We note that%
\begin{align}
\mathcal{L}_{m}^{\prime}\big|_{g}  & =\frac{\partial\mathcal{L}_{m}}%
{\partial\tilde{g}^{\mu\nu}}\frac{\partial(e^{\varphi}g^{\mu\nu})}%
{\partial\varphi}-W^{\prime}=\tilde{g}^{\mu\nu}\frac{\partial\mathcal{L}_{m}%
}{\partial\tilde{g}^{\mu\nu}}-W^{\prime}\nonumber\\
& =\left(  \frac{1}{2}\tilde{T}+2\mathcal{L}_{m}\right)  -W^{\prime}=\left(
\frac{1}{2}T+2\mathcal{L}_{m}\right)  -W^{\prime}\label{23}%
\end{align}

where we have used $\tilde{T}_{\mu\nu}=2\frac{\partial\mathcal{L}_{m}%
}{\partial\tilde{g}^{\mu\nu}}-\tilde{g}_{\mu\nu}\mathcal{L}_{m}$ and
$\tilde{T}=2\tilde{g}^{\mu\nu}\frac{\partial\mathcal{L}_{m}}{\partial\tilde
{g}^{\mu\nu}}-4\mathcal{L}_{m}$. From (\ref{19}) $\tilde{h}=e^{2\varphi}h$,
which when combined with (\ref{21}) yields $\tilde{T}=T$. Therefore, the
consistency condition (\ref{16}), or equivalently (\ref{22}), is satisfied, in
general, for any matter lagrangian (that does not depend upon derivatives of
the metric). We conclude that the consistency conditions (\ref{16}) and
(\ref{17}) hold, in general, so that the translation of the JF EoM, derived
from the JF action (\ref{1a}), under the conformal transformation (\ref{2}),
will coincide with the EF EoM derived from the EF action (\ref{1b}).

\section{Summary}

\ \ We have considered a scalar-tensor theory of a general form given by
(\ref{1a}) and (\ref{1b}). The Jordan frame and Einstein frame actions are
equivalent, and a complete set of equations of motion can be derived from the
action in each frame. Using the conformal transformation, the set of equations
of motion can be transcribed from the JF to the EF, i.e., the JF equations of
motion are written in terms of EF variables. It was shown that the JF field
equations for the matter fields map onto the EF field equations\ for those
fields, and that the JF Einstein equation and JF dilaton equation map onto the
corresponding EF equations, provided that consistency conditions, given by
(\ref{16}) and (\ref{17}), are satisfied. It was shown that these conditions
are always satisfied, so that a solution set obtained in one conformal frame
can be translated, with confidence, via the conformal transformation, to the
other frame.

\appendix{}

\section{}

\ \ Here we list some useful relations for the translation of equations of
motion. We start by writing the Jordan frame and Einstein frame covariant
derivatives, respectively,%
\begin{equation}
\tilde{\nabla}_{\mu}B_{\nu}=\partial_{\mu}B_{\nu}-\tilde{\Gamma}_{\mu\nu
}^{\lambda}B_{\lambda},\ \ \ \ \ \nabla_{\mu}B_{\nu}=\partial_{\mu}B_{\nu
}-\Gamma_{\mu\nu}^{\lambda}B_{\lambda}\label{a1}%
\end{equation}

with $B_{\nu}=\partial_{\nu}\varphi$. The JF connection $\tilde{\Gamma}%
_{\mu\nu}^{\lambda}$, under the conformal transformation $\tilde{g}_{\mu\nu
}=\Omega^{2}g_{\mu\nu}$, where $\Omega=F^{-1/2}=e^{-\varphi/2}$, is related to
the EF connection $\Gamma_{\mu\nu}^{\lambda}$ by \cite{BDbook},\cite{Kaiser}%
\begin{align}
\tilde{\Gamma}_{\mu\nu}^{\lambda}  &  =\Gamma_{\mu\nu}^{\lambda}+\Omega
^{-1}\left[  \delta_{\mu}^{\lambda}\partial_{\nu}\Omega+\delta_{\nu}^{\lambda
}\partial_{\mu}\Omega-g_{\mu\nu}\partial^{\lambda}\Omega\right] \nonumber\\
&  =\Gamma_{\mu\nu}^{\lambda}-\frac{1}{2}\left[  \delta_{\mu}^{\lambda
}\partial_{\nu}\varphi+\delta_{\nu}^{\lambda}\partial_{\mu}\varphi-g_{\mu\nu
}\partial^{\lambda}\varphi\right]  \label{a2}%
\end{align}

Then (\ref{a1}) and (\ref{a2}) give%
\begin{align}
\tilde{\nabla}_{\mu}\partial_{\nu}\varphi &  =\partial_{\mu}\partial_{\nu
}\varphi-\Gamma_{\mu\nu}^{\lambda}\partial_{\lambda}\varphi+\frac{1}{2}\left[
\delta_{\mu}^{\lambda}\partial_{\nu}\varphi+\delta_{\nu}^{\lambda}%
\partial_{\mu}\varphi-g_{\mu\nu}\partial^{\lambda}\varphi\right]
\partial_{\lambda}\varphi\nonumber\\
&  =\left(  \partial_{\mu}\partial_{\nu}\varphi-\Gamma_{\mu\nu}^{\lambda
}\partial_{\lambda}\varphi\right)  +\frac{1}{2}\left[  2\partial_{\mu}%
\varphi\partial_{\nu}\varphi-g_{\mu\nu}\partial^{\lambda}\varphi
\partial_{\lambda}\varphi\right]  \nonumber\\
&  =\nabla_{\mu}\partial_{\nu}\varphi+\partial_{\mu}\varphi\partial_{\nu
}\varphi-\frac{1}{2}g_{\mu\nu}(\partial\varphi)^{2}\label{a3}%
\end{align}

We also have%
\begin{equation}
\tilde{\square}\varphi=\frac{1}{\sqrt{\tilde{g}}}\partial_{\mu}\left[
\sqrt{\tilde{g}}\tilde{g}^{\mu\nu}\partial_{\nu}\varphi\right]  =e^{\varphi
}\left[  \square\varphi-(\partial\varphi)^{2}\right]  ,\ \ \ \ \ (\tilde
{\partial}\varphi)^{2}=\tilde{g}^{\mu\nu}\partial_{\mu}\varphi\partial_{\nu
}\varphi=e^{\varphi}(\partial\varphi)^{2}\label{a4}%
\end{equation}

Also%
\begin{align}
\tilde{R}_{\mu\nu}  & =R_{\mu\nu}+2\nabla_{\mu}\partial_{\nu}\ln\Omega
+g_{\mu\nu}g^{\alpha\beta}\nabla_{\alpha}\partial_{\beta}\ln\Omega\nonumber\\
& -2(\partial_{\mu}\ln\Omega)\partial_{\nu}\ln\Omega+2g_{\mu\nu}g^{\alpha
\beta}(\partial_{\alpha}\ln\Omega)\partial_{\beta}\ln\Omega\label{a5}%
\end{align}

or, since $\ln\Omega=-\frac{1}{2}\varphi$,%
\begin{equation}
\tilde{R}_{\mu\nu}=R_{\mu\nu}-\nabla_{\mu}\partial_{\nu}\varphi-\frac{1}%
{2}g_{\mu\nu}\square\varphi-\frac{1}{2}\partial_{\mu}\varphi\partial_{\nu
}\varphi+\frac{1}{2}g_{\mu\nu}(\partial\varphi)^{2}\label{a6}%
\end{equation}

which leads to%
\begin{equation}
\tilde{G}_{\mu\nu}=G_{\mu\nu}-\nabla_{\mu}\partial_{\nu}\varphi+g_{\mu\nu
}\square\varphi-\frac{1}{2}\partial_{\mu}\varphi\partial_{\nu}\varphi-\frac
{1}{4}g_{\mu\nu}(\partial\varphi)^{2}\label{a7}%
\end{equation}

\end{document}